\documentclass{revtex4}
\usepackage{epsf,graphicx}

\begin{document}
\title{On the fragmentation of multiply charged sodium clusters}
\author{H.I. Hidmi $^1$, D.H.E. Gross $^2$, and H.R. Jaqaman $^1$}

\address{$^1$ Bethlehem University, Bethlehem, Palestine \\
$^2$ Hahn-Meitner-Institut
Berlin, Bereich Theoretische Physik, \\14109 Berlin, Germany\\
and\\
Freie Universit\"at Berlin}
\date{\today}

\begin{abstract}
The fragmentation of multiply charged atomic sodium clusters of mass 200 is 
investigated using the Micro-canonical Metropolis Monte Carlo (MMMC) 
statistical technique for excitation energies up to 200 eV and for cluster 
charges up to $+9e$.
In this work we present caloric curves and charged and uncharged fragment 
mass distributions for clusters with charges 0, 2, and 4. The caloric curves 
show a dip at the critical point implying a negative specific heat, as 
expected for finite systems, while the fragment mass distributions 
corroborate the picture of a phase transition from one dominant liquid-like 
cluster
to complete vaporization.
\end{abstract}
\maketitle
\section{\bf Introduction}
The thermo-statistical investigation of atomic clusters is interesting for 
several reasons. On the practical side,
the study of mesoscopic many-body systems is becoming increasingly important 
in modern technology which is very quickly approaching the nano scale where 
quantum and finite size effects begin to play a role. On the theoretical 
side, the study of small systems also gives a rich and deep insight into the 
fundamental concepts of statistical mechanics
and tries to answer some important questions \cite{gross180,gross182}
such as how large must a system be in order to show phase transitions, or 
what is the microscopic
origin of the Second Law of Thermodynamics, and does it hold in small 
systems?
Here we hope to contribute to answering the first question.

Phase transitions are identified by the shape of the caloric
curve, the relation between the temperature and the excitation energy
of the system.  In macroscopic physics it is usually assumed that the system 
approaches the thermodynamic limit
where the number of particles in the system $N\to\infty$ and a phase 
transition is
characterized by the fact that the temperature at the phase transition 
remains constant while
energy is supplied to the system. In this limit, surface effects ($\propto 
N^{2/3})$ can be neglected relative to the
leading bulk quantities ($\propto N$).

  Dealing with finite isolated
systems, the picture becomes different. As the
starting point of a conventional canonical description is the
assumption of a homogeneous density distribution in the thermodynamic
limit, the canonical description is not suited for the description of
phase transitions in small clusters.  The main physical effect of a
phase transition of first order is the creation of inhomogeneities
with a separation of two coexisting phases by interfaces.  Here
surface effects are important and
the signal of a first order phase transition is seen as a dip in the
caloric curve (temperature versus energy) at the critical temperature,
as shown in Figure \ref{fig41}.  This is
similar to observations on small isolated nuclear systems
\cite{chbihi95,dAgostino00,gross00} and was predicted by Gross {\em et 
al.}\cite{gross72}.
This indicates that the heat capacity is not a
positive definite quantity any more and can even take negative values.

This behavior can only be obtained in a micro-canonical description
which takes into account the strict conservation of mass, charge, and
energy, and allows for a proper treatment of surface effects.
In this approach, the phase transition can be
attributed to the opening of new decay
channels, i.e, the population of additional regions of the phase space and 
the
concomitant increase in the density of phase space, i.e. in the number 
$W(E_{tot})$ of micro-states
of the system with total energy ${E_{tot}}$. The
entropy of the system is related to the phase space density through 
Boltzmann's Principle
${S(E_{tot})={k\ln W(E_{tot})}}$, where $k$ is Boltzmann's constant.
The thermodynamic temperature of the system can then be obtained from the 
entropy by the relation:
\begin{eqnarray}
{{1\over{T_{ther}}}\equiv{\beta(E_{tot})}={{\partial{S(E_{tot})}}\over{\partial{E_{tot}}}}}
\end{eqnarray}

In the present work the fragmentation of charged atomic Na
clusters of mass 200, with cluster charges ranging from 0 to 9, is 
investigated
assuming that statistical equilibration is achieved at some point before
the fragmentation process takes place. The statistical method used is the 
Micro-canonical
Metropolis Monte Carlo (MMMC) technique, described in detail
in \cite{gross90,zhang,gross95,olga,gross151,olga3}. The fundamental 
assumption is that, at
equilibrium, all micro-canonically accessible phase-space cells
$\xi_i$ corresponding to a fixed energy (and any other conserved quantities) 
are equally probable.
The computational simulation of the fragmentation of a sodium cluster is 
performed
as follows: N atoms (200  in our case) are combined randomly into fragments 
(or subclusters)
which are placed randomly inside a spherical volume
of specified radius ($r\times N^{1/3}$) where $r$ is the system radius 
parameter
that is taken  to be  $6 \AA$ in this work. Other
values of $r$ were also used and the results were found to depend weakly on 
the choice of $r$.
Thus we assume that fragments explore
all the accessible phase space when moving within a well defined
freeze-out volume. This scenario is different from that where
the fragments are in isobaric equilibrium with a pressure defined by
the evaporation rate as assumed by Schmidt et.al. \cite{Labastie}.

The system is given a total charge $Z$, and a given excitation
energy or heat. The charge and excitation energy are distributed randomly 
among the various fragments.
The resulting fragment configuration corresponds to a point $\xi_i$ in the 
reduced phase space of the system
and the weight of the phase space at that point is calculated.
So, starting from this initial phase-space point, with weight $W_i$, one 
goes to a
neighboring point $\xi_f$ in phase space by, for example,  repositioning one 
of the fragments,
by combining the masses and charges of two fragments or by splitting the 
mass and
charge of an existing fragment into two fragments and repositioning them. 
One then
calculates the weight $W_f$ of the new point in the reduced phase space, and 
the new state
is accepted with a probability $\Pi=W_f/W_i$. It is worth mentioning that 
the
weights are calculated using detailed balance. The states are not
generated by the computer with an equal probability, but rather with an 
"apriori-
probability" which depends on the way we sample the fragments. This will 
give
different values for the transition probability from state $1$ to state $2$, 
and
the transition from state $2$ to state $1$. So, one has to correct for
this transition probability difference in order to keep detailed balance 
between
the initial and final states. The correct inclusion of detailed balance for 
charged
clusters is an improvement over the earlier work \cite{olga,gross151,olga3} 
where detailed
balance was included only approximately for charged clusters.

Clusters of mass 200 atoms and charges ranging from zero up to 9 were 
studied.
It was found that for charges $\geq 5$ the
calculation becomes unstable and this was observed
as irreproducible fluctuations in the caloric curve that vary chaotically if 
the
calculations are repeated with a different random number generator
or with a different number of phase space points (typically a few millions)
sampled at each energy. This puzzling behavior
is due to the fact that as the charge of a fixed-mass cluster
is increased, the long-range Coulomb repulsive force overwhelms the much
weaker short-range van der Waals attractive forces with the result that
the cluster becomes unstable even at zero excitation energy. Such a 
situation has
been observed experimentally by Chandezon {\em et al.} \cite{chand1,chand2} 
who
investigated the critical sizes necessary for the production
of multiply charged sodium clusters and
found that for a charge of 5, the smallest mass of the cluster would be
$200\pm5$ atoms, a result slightly dependent on the method used for 
producing
and ionizing the cluster. For a total cluster charge of 4 the critical size 
is about 120.
Charged clusters smaller than the critical size where found to decay
immediately by emitting a light singly charged
fragment and thus cannot achieve thermal equilibration. This behavior 
contrasts
with the nuclear case where the Coulomb force has to compete with
the much stronger nuclear force and where $Z$ usually takes values
close to half the number of nucleons.

\section{\bf RESULTS}

\begin{center}
\begin{figure}
\includegraphics*[bb =88 25 590 735, angle=-90, width=12cm,
clip=true]{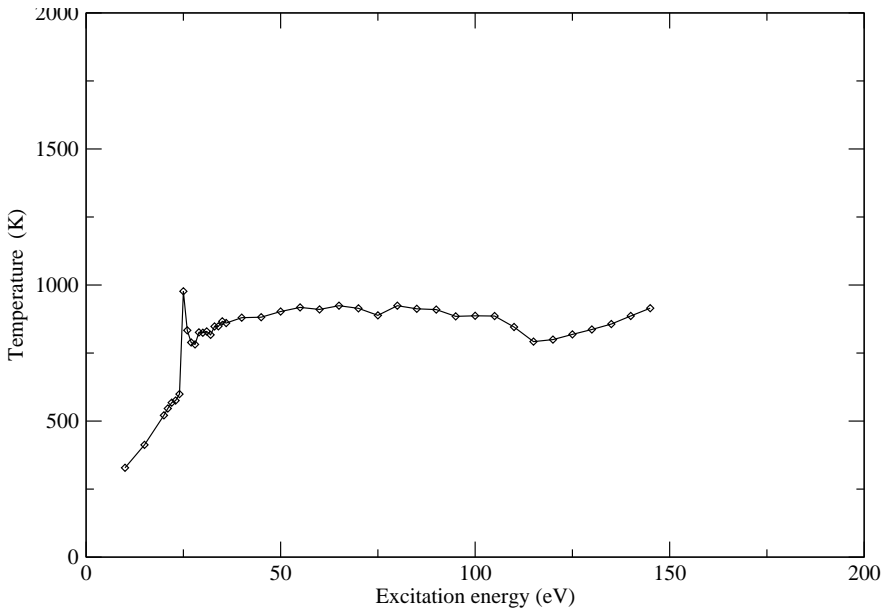}
\caption{Caloric curve for Na clusters of mass 200 atoms with total 
charge$=4e$.
The temperature dip at E about 110 eV is the signature of the phase 
transition in small systems.
This curve is obtained with statistics of 10 $\times 10^6$ events per energy 
step.\label{fig41}}
\end{figure}
\end{center}

Figure \ref{fig41} shows the caloric curve, or the thermodynamic temperature 
as a function of
the excitation energy, that we obtained in our MMMC
simulation for a cluster of sodium with a mass of 200 atoms and a charge of 
$+4e$. It is
clear from the figure that the temperature of the system increases
sharply with energy for excitation energies below 30 eV (or 0.15eV/atom), 
then it reaches a plateau
at a temperature of about 900K. At energies above 100 eV the cluster starts
to evaporate rapidly, its size shrinks appreciably (see also Figure 
\ref{fig5} below)
and its temperature is reduced by about 100K. This
is indicated by the dip in the temperature curve at about 110 eV excitation
energy. It corresponds to a negative specific heat capacity of a
value of about $-0.45 cal/g K$.
Similar results were obtained for clusters of
charges from 0 to 3, with the dips at the end of the phase transition 
plateau becoming deeper as the charge is reduced.
The negative heat capacity is observed in small systems, in contrast
to infinite system in which the heat capacity remains constant during phase 
transition.

\begin{center}
\begin{figure}
\includegraphics*[bb =88 25 590 735, angle=-90, width=12cm,
clip=true]{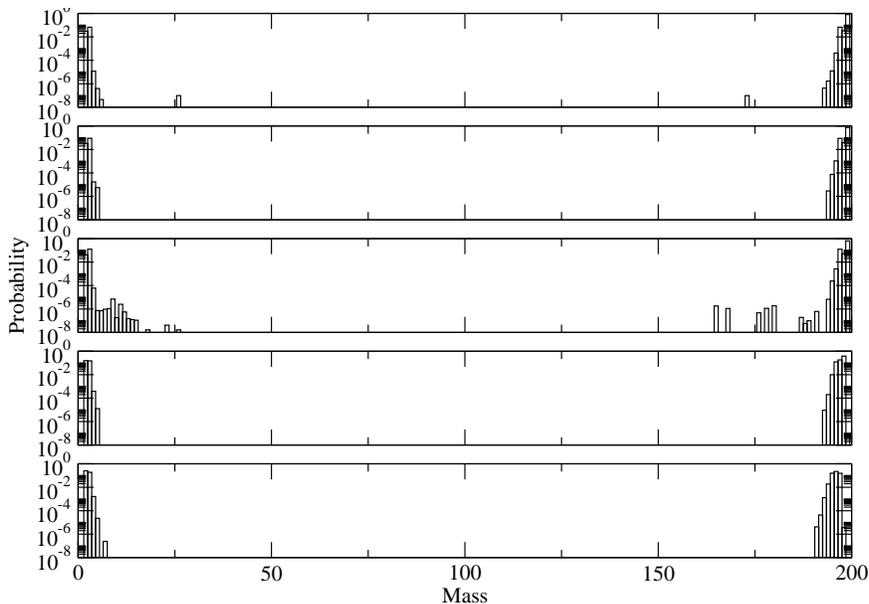}
\caption{Mass distribution for Na clusters of 200 atoms and a cluster charge 
of 2e near the peak
of the caloric curve at 25 eV. The panels correspond to excitation energies 
of 23, 24, 25, 26,
and 27 eV from top to bottom.
\label{figpeak}}
\end{figure}
\end{center}

The puzzling peak at $25 eV$ excitation energy in the
caloric curve is reproducible and independent of the calculation. It has 
been verified
with very high statistics, $1\times 10^8$ events per energy
step, and with small energy steps of $1 eV$. This peak was also obtained by 
Gross {\em et al.} \cite{olga}
who carried out similar calculations that however did include detailed 
balance only in
average manner.
The mass distribution of the cluster around the peak is shown in Figure 
\ref{figpeak} for
the case of clusters with charge 2.  The mass distribution at the peak,
is found to have convergence problems, and to have uncertainties on
the order of $1\times 10^{-6}$.

\begin{center}
\begin{figure}
\includegraphics*[bb =67 25 590 735, angle=-90, width=12cm,
clip=true]{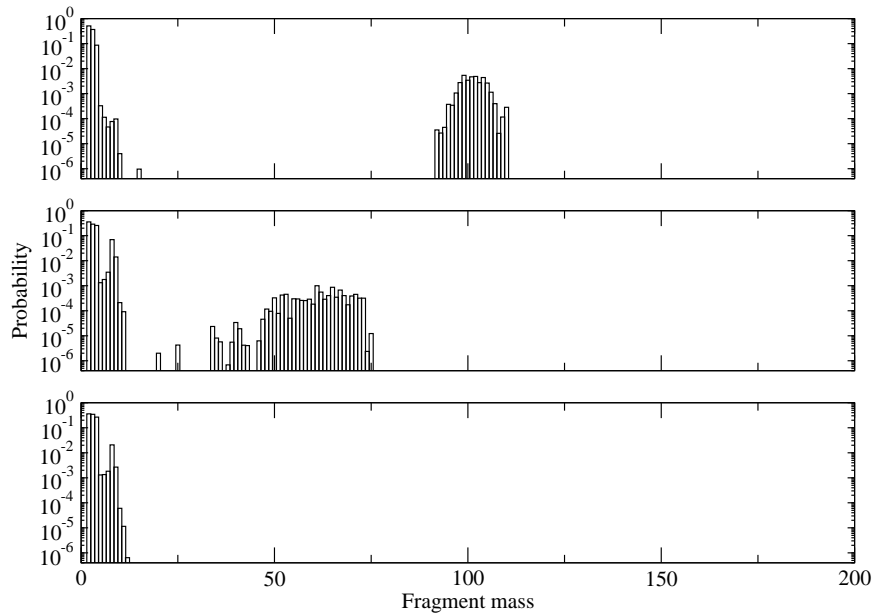}
\caption{The mass distribution for Na clusters of mass 200 atoms and zero 
charge at excitation energies of
90, 110, 130 eV from top to bottom.\label{fig02}}
\end{figure}
\end{center}

Figure \ref{fig02} shows the mass distribution for the fragmentation of 
uncharged
Na clusters of mass 200 at three different excitation
energies: before ($E=90 eV$), in the middle of
($E=110 eV$), and right after ($E=130 eV$) the  temperature dip . This 
figure
excludes the abundantly produced neutral monomers. We see from this figure 
that at all energies the dimers and
trimers occur with highest probability. This figure shows how the 
fragmentation process
takes place. For energies $\leq 90 eV$ large fragments containing more than
100 atoms coexist with very light fragments. At the dip energy of 110 eV we 
observe a broad distribution of fragments
of sizes between 25 and 75 atoms, with probabilities 2-3 orders of magnitude 
smaller
than those of the light fragments. As the excitation energy is increased to 
$130 eV$
only small fragments consisting of 10 atoms or less survive, with dimers, 
trimers, quadrimers
and octamers singled out with relatively higher probabilities than the other 
multimers, in
agreement with the shell model.

\begin{center}
\begin{figure}
\includegraphics*[bb =88 25 590 735, angle=-90, width=12cm,
clip=true]{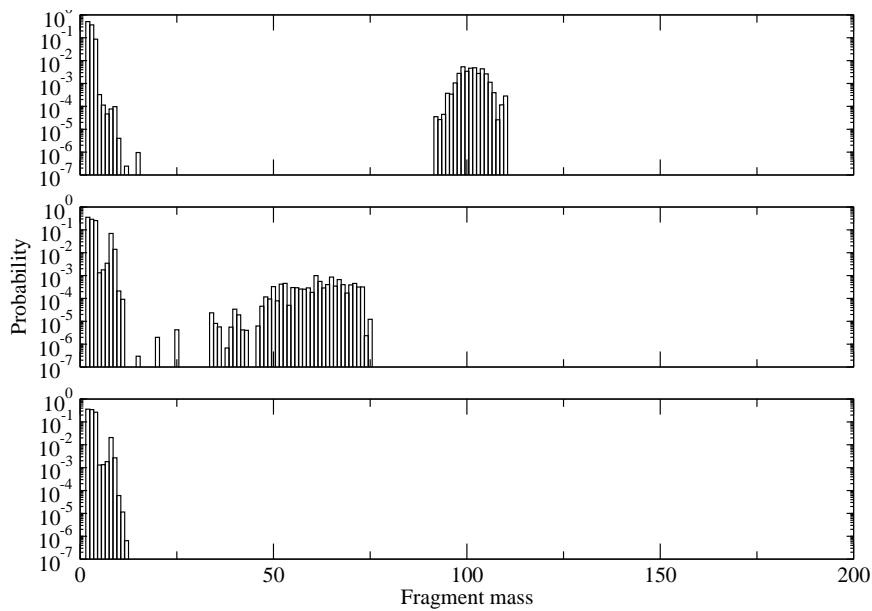}
\caption{Total mass distribution, excluding the neutral monomers,
in the fragmentation of doubly charged Na clusters of mass 200 at
energies 90, 110, 130 eV from top to bottom.
\label{fig22}}
\end{figure}
\end{center}

\begin{center}
\begin{figure}
\includegraphics*[bb =67 25 590 735, angle=-90, width=12cm,
clip=true]{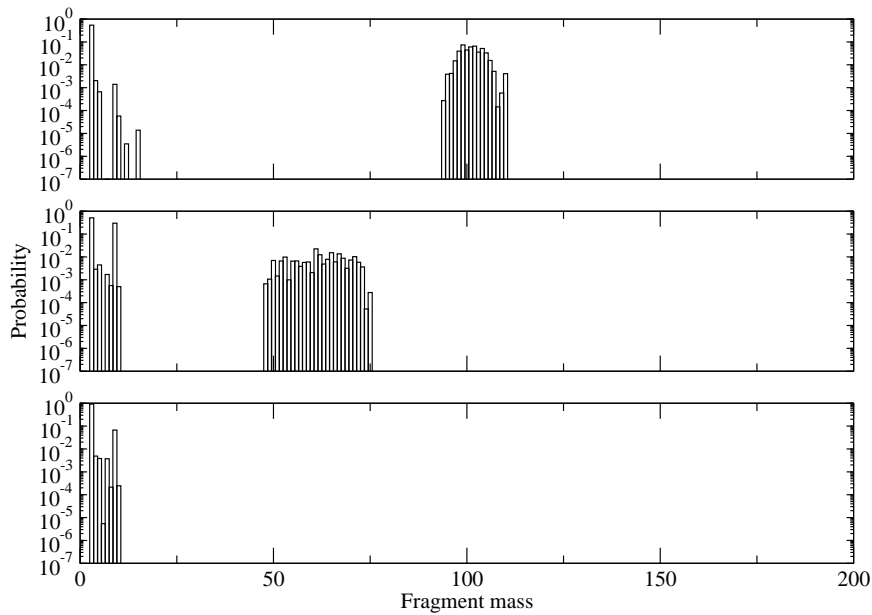}
\caption{The mass distribution for charged fragments resulting from the 
fragmentation
of doubly charged Na clusters of mass
200 at energies of 90, 110, 130 eV from top to bottom. All fragments are 
singly charged.
.\label{fig23}}
\end{figure}
\end{center}

In Figure \ref{fig22} we plot the total mass distribution for Na 200 and a 
cluster charge
of 2 at the three different excitation energies of 90, 110, and 130 eV, from 
top to bottom. This
figure again excludes the neutral monomers, but includes neutral and charged 
multimers.
The total mass distribution for charge 2 is very similar to that shown in
Figure \ref{fig02} for the zero charge case. It is also instructive to 
compare Figure \ref{fig22}
with Figure \ref{fig23}, which gives the mass distribution of the charged 
fragments
for the same system (charge 2). Only singly charged fragments are observed 
at the energies considered in this figure
and the heavy fragments at the excitation energies of 90 and 110 eV
have a higher  probability than in the total mass distribution or in the 
neutral case.
It is also noted that among the charged fragments trimers
and 9-mers are present with higher probability than the others. No dimers, 
and very few
octamers were observed as expected from shell effects, namely, neutral 
dimers
and octamers are favored, as well as singly charged trimers and 9-mers are 
favored. The
dominance of singly-charged trimers for such lightly charged fragments
agrees with experimental evidence presented
by C. Guet {\em et al.} \cite{cguet} who contrast this with the sudden 
Coulomb explosion
of highly charged clusters in which singly charged monomers dominate. These 
latter
events cannot be generated in our statistical model because they occur on a 
shorter time
scale than that needed for the establishment of thermal equilibration.

\begin{center}
\begin{figure}
\includegraphics*[bb =67 25 590 735, angle=-90, width=12cm,
clip=true]{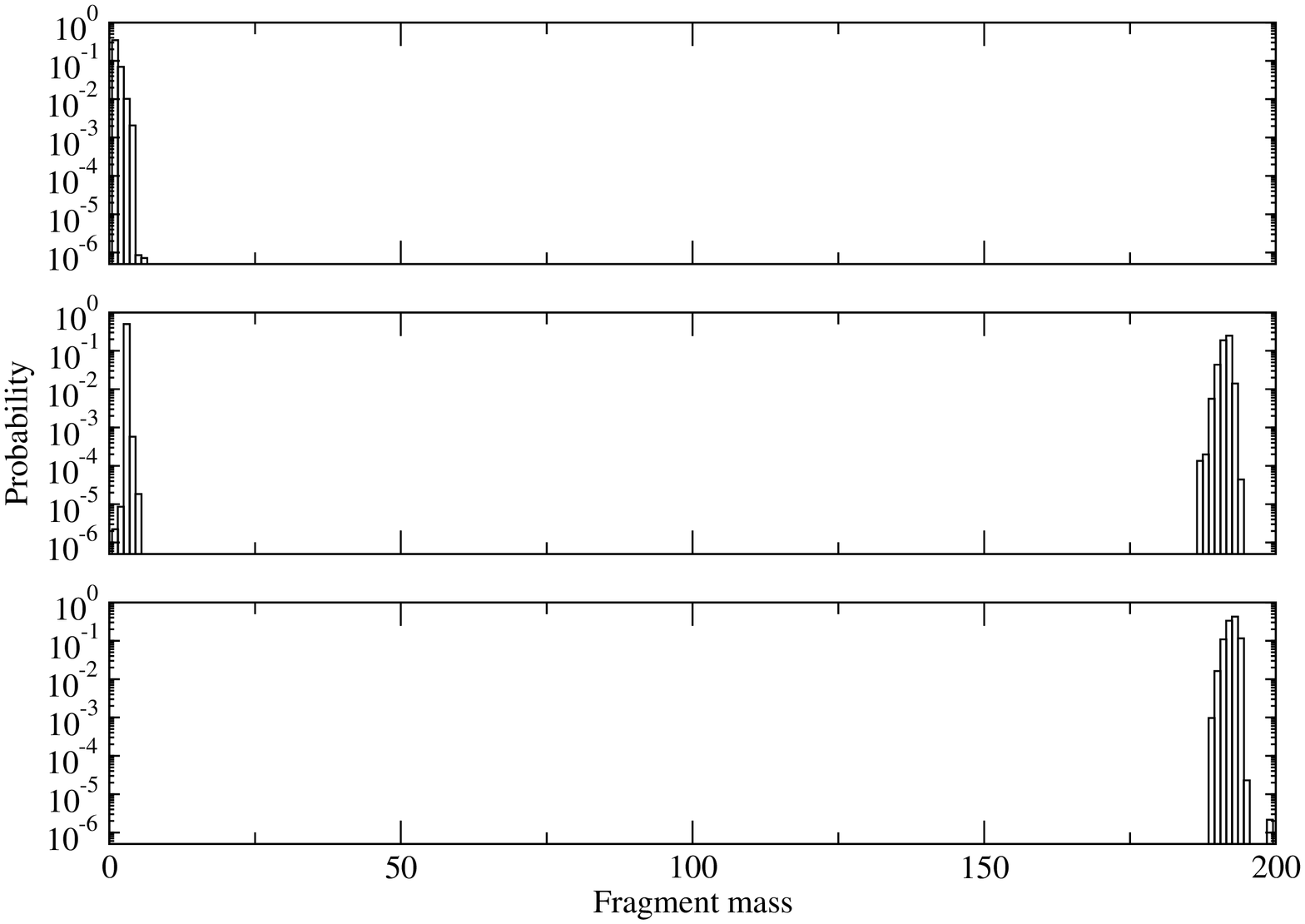}
\caption{Mass distribution of charged fragments from the fragmentation of Na 
clusters of mass 200
and a total charge $=2e$ at an
excitation energy of 30 eV. The plots are for fragments with charge 0, 1, 
and 2 from top to
bottom.
\label{fig24}}
\end{figure}
\end{center}

Figure \ref{fig24} shows the mass distribution of fragments resulting from 
the
fragmentation of doubly charged Na clusters of mass 200 at an excitation 
energy of $30 eV$. The top
panel shows the distribution of neutral fragments while the middle and 
bottom panels show
the distribution of singly and doubly charged fragments, respectively.
Note that for the singly charged fragments, trimers occur with highest 
probability among
other fragments, as expected from shell effects. It is also clear from this 
figure that, at
this relatively low excitation energy, the system fragments by emitting a 
few light fragments
(of mass less than 5) and leaving a heavy residual fragment of about 195 
atoms. It is also noted that
there are no doubly charged light fragments and no neutral heavy fragments. 
Because of the
competition between Coulomb and surface effects, the charges prefer to 
reside
on the heavy  fragments.

\begin{figure}[htbp]
\begin{center}
\includegraphics*[bb =67 25 590 735, angle=-90, 
width=12cm,clip=true]{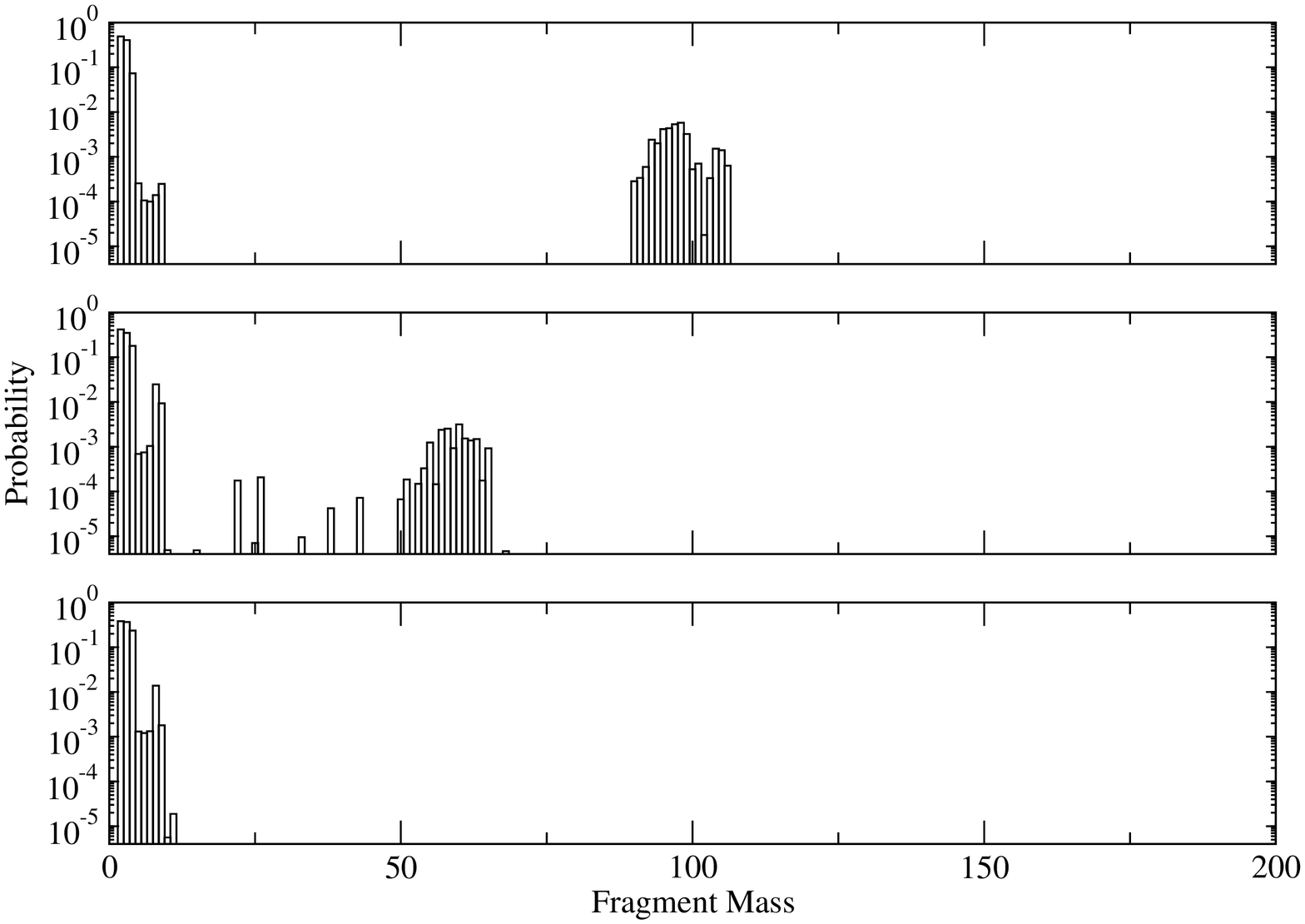}
\caption{ Total mass distribution (excluding the neutral monomers) for Na
cluster of mass 200 and  total charge$=4e$. The plots are taken at energies 
of 90, 110, and 130 eV from top to bottom.
\label{fig42}}
\end{center}
\end{figure}

Figure \ref{fig42} shows the total mass distribution, excluding the neutral
monomers, for the fragmentation of clusters of mass 200 atoms and a cluster 
charge of 4 at 90, 110 1nd 130 eV.
The excitation energies chosen are the same as those previously used for 
neutral and doubly charged clusters,
and the results are very similar to those in Figure \ref{fig02} and Figure 
\ref{fig22}.
The only notable difference is that the distribution of the heavy fragments 
for the charge 4 case at 110 eV
has a narrower range than that for charge 0 or 2 cases at the same 
excitation energy.
This may reflect a slight shift in the location of the temperature dip 
indicating
that increasing the charge results in a corresponding modification in the 
properties of the phase transition.


\begin{center}
\begin{figure}
\includegraphics*[bb =67 25 590 735, angle=-90, width=12cm,
clip=true]{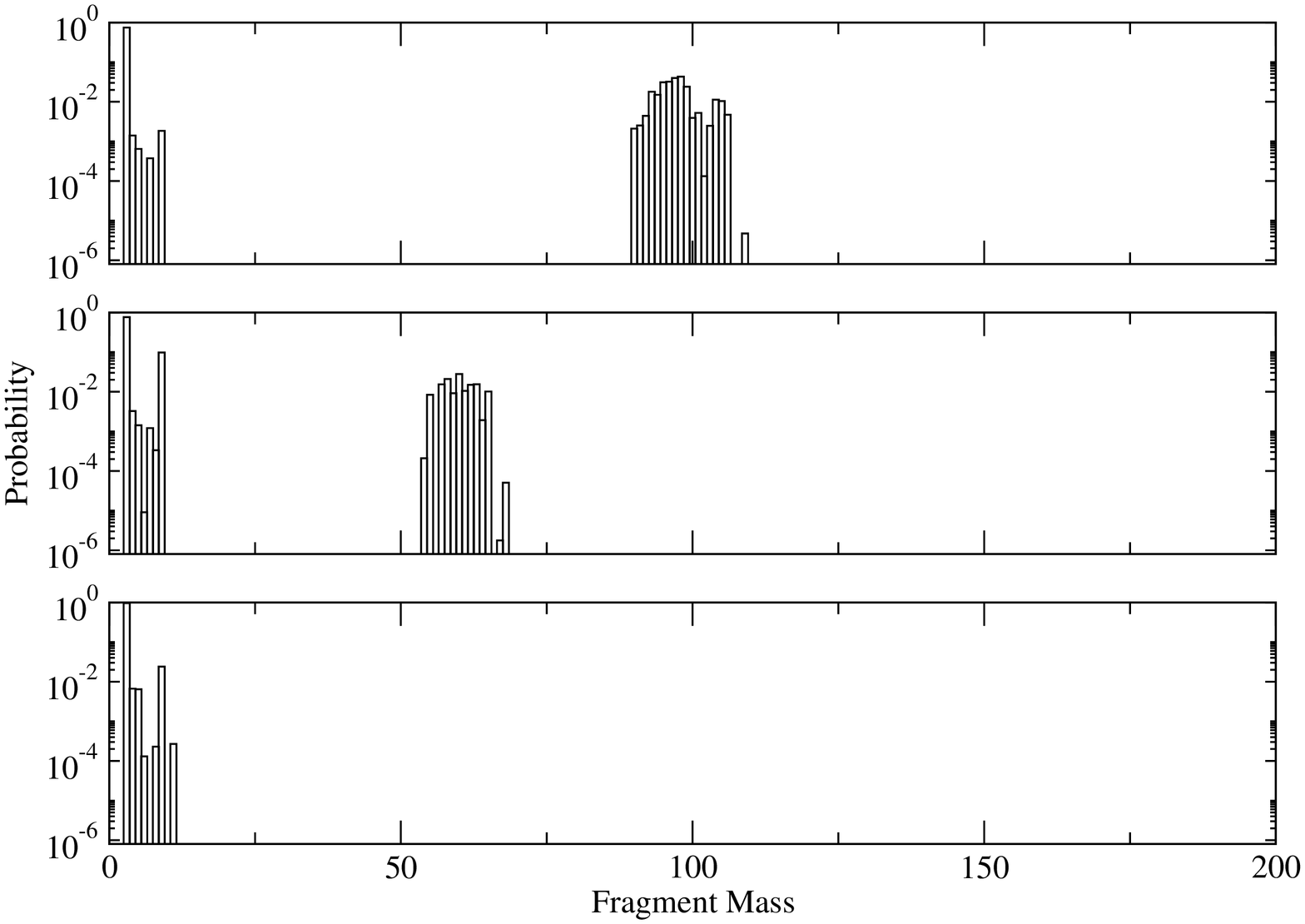}
\caption{Mass distribution of charged fragments for Na clusters of mass 200
and with total charge$=4e$. All the fragments are singly charged and the 
plots are taken at excitation
energies of 90, 110, and 130 eV from top to  bottom.}
\label{fig43}
\end{figure}
\end{center}

In figure \ref{fig43} we present the mass distribution of the charged 
fragments, for a
cluster of mass 200 and a total charge of 4 at energies
90, 110, and 130 eV from top to bottom. This
figure also shows the fragmentation of the cluster as a function of energy 
around the
negative heat capacity region. All fragments at these energies are singly 
charged.
In fact we did not observe multiply charged fragments for
excitation energies above 30 eV.
It is clear from this figure that at these energies, the singly charged 
trimers are the most
dominant among all others, followed by the singly charged 9-mers. This is in 
agreement with
the electronic shell effect, which predicts that singly charged trimers and 
9-mers are
strongly bound (with full energy levels) and have relatively high 
dissociation energy.
Note also in this figure that there is no charged dimers, again in agreement 
with
the shell model.
Both fission and evaporation compete in the fragmentation process, and the 
end
result of these processes is that fragments of smaller size are produced as 
the energy is
increased. At energies $\geq 130 eV$ only fragments with mass less than 10
are observed.

\begin{center}
\begin{figure}
\includegraphics*[bb =82 25 590 735, angle=-90, width=12cm,
clip=true]{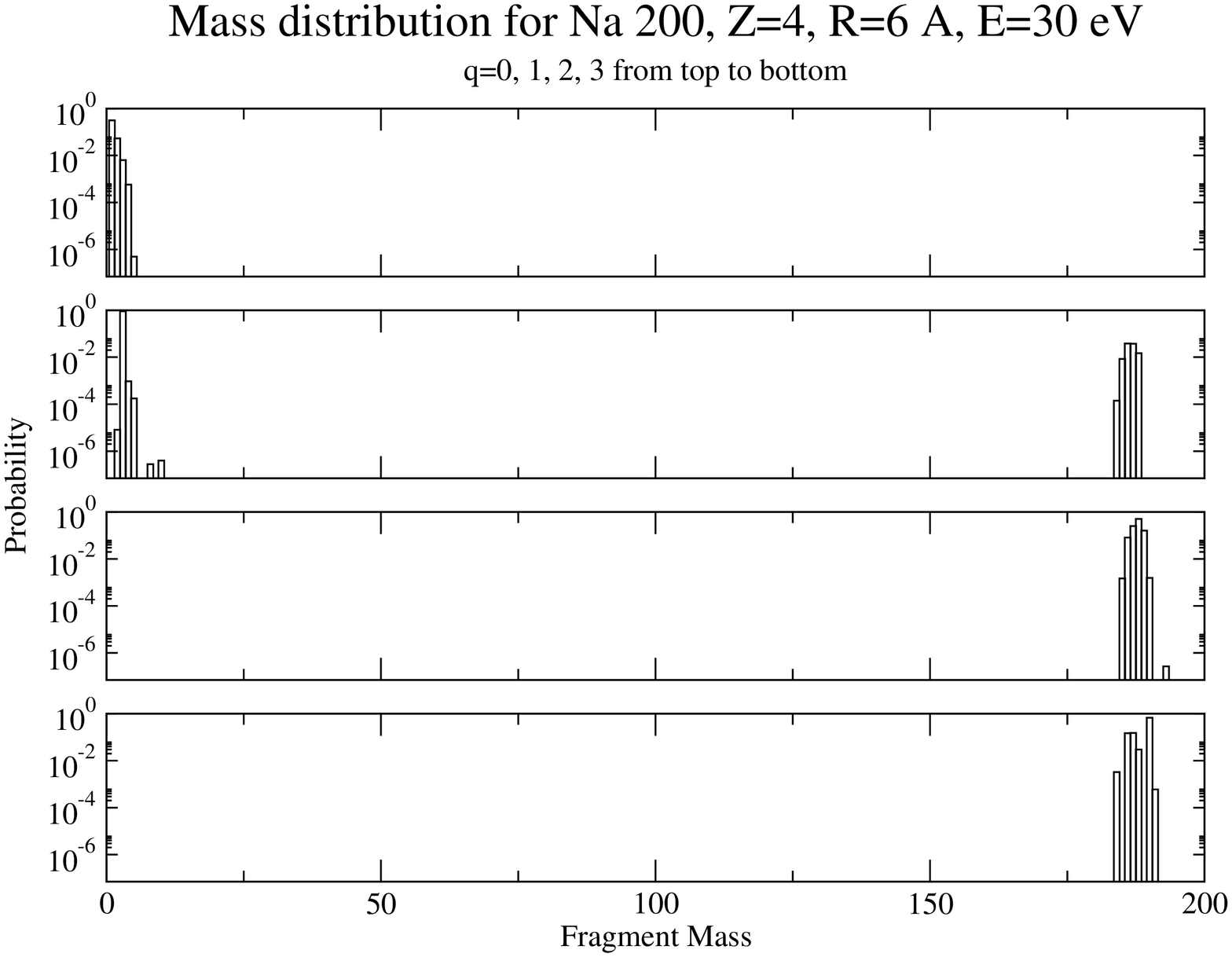}
\caption{Mass distribution of fragments, sorted by their charge, in the 
fragmentation of Na clusters of mass 200
atoms and with total charge$=4e$ at an
excitation energy of 30 eV. The plots are for fragments with charge 0, 1, 2, 
and 3 from top to bottom, respectively.
\label{fig44}}
\end{figure}
\end{center}

Figure \ref{fig44} gives the mass distribution of the fragments sorted 
according to their
charge at a total excitation energy of $30 eV$.
At this energy it is possible to observe multiply charged fragments. Light 
fragments containing
less than 10 atoms are either uncharged or singly charged
while the heavy fragments with masses in the range of
182-192 atoms are all charged up to a maximum charge of 3.
Note that in this case there are no fragments which
carry the whole charge 4, in contrast to the Z=2 case. This is due to the 
fact
that at the beginning of the excitation process, a singly charged trimer
is emitted from the cluster leaving a big fragment with a maximum charge of 
3.
It is also clear from this figure that the singly charged
trimers are the most probable species among all others.
Above 30 eV only singly charged fragments can exist, and the fragment size
is decreased as the excitation energy is increased.

The development of the fragmentation process as the excitation energy is 
increased can
be seen from Fig.\ref{fig5} which gives the two largest
fragments as a function of excitation energy for clusters with total charges 
of 0, 2, and 4.
In the three cases the cluster remains
as a big piece while a few trimers and quadrimers are evaporated up to an 
excitation energy of $110 eV$.
Of course monomers and dimers exist, and some fragments with masses upto 8 
can be
found but with small probability, as can be seen by comparing with Figure 
\ref{fig44}. The three panels
for the different charges look very similar.
Above an excitation energy of $110 eV$,
only fragments with mass less than 10 can exist, as can also be seen by 
comparing with
Figures \ref{fig02}, \ref{fig22} and \ref{fig42}.

\begin{center}
\begin{figure}
\includegraphics*[bb =82 25 590 735, angle=-90, width=12cm,
clip=true]{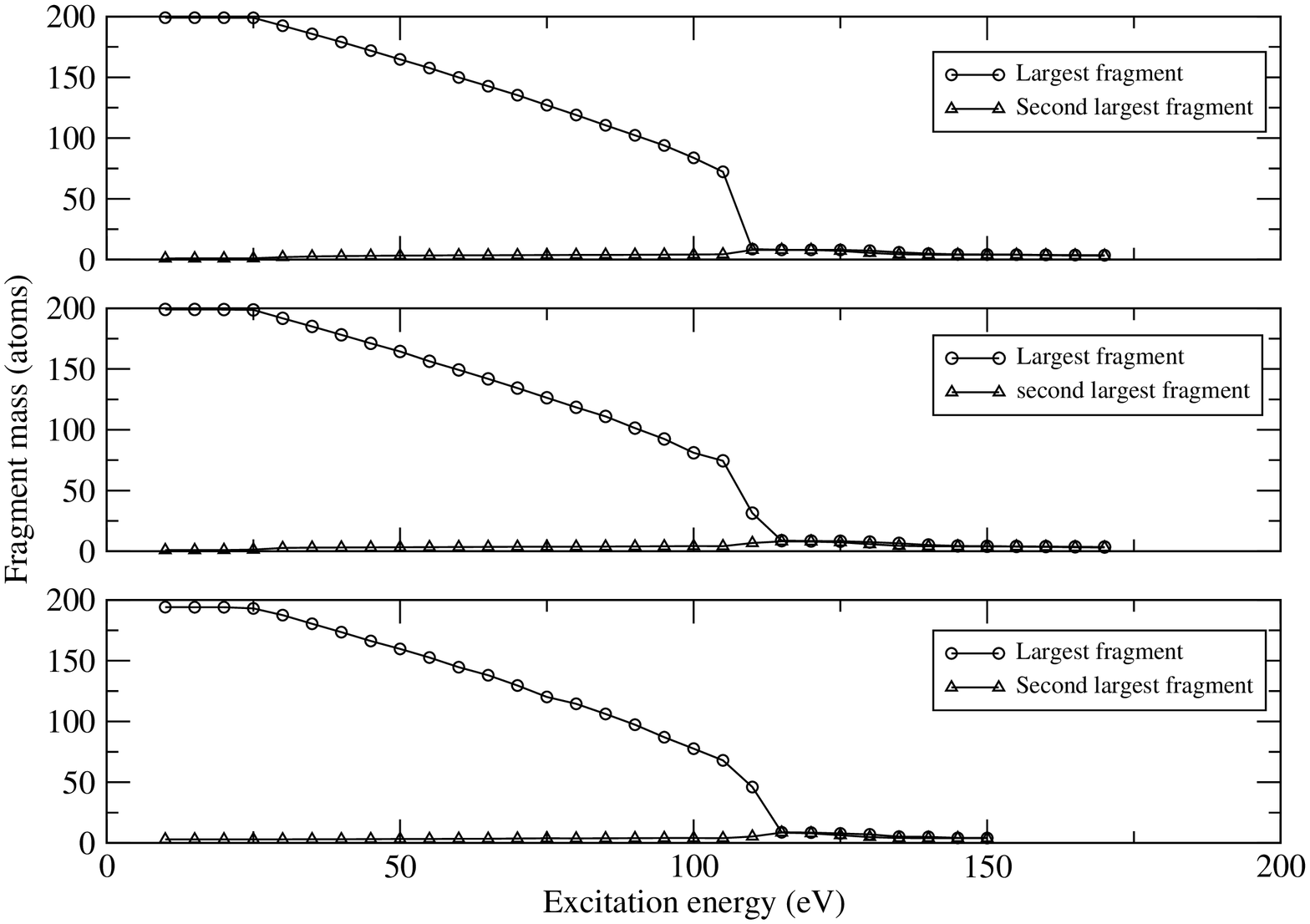}
\caption{The two largest fragments resulting from the fragmentation of Na
clusters of mass 200 and with total charges of 0, 2e, and 4e, from top to 
bottom, respectively.
\label{fig5}}
\end{figure}
\end{center}

\begin{center}
\begin{figure}
\includegraphics*[bb =82 25 590 735, angle=-90, width=12cm,
clip=true]{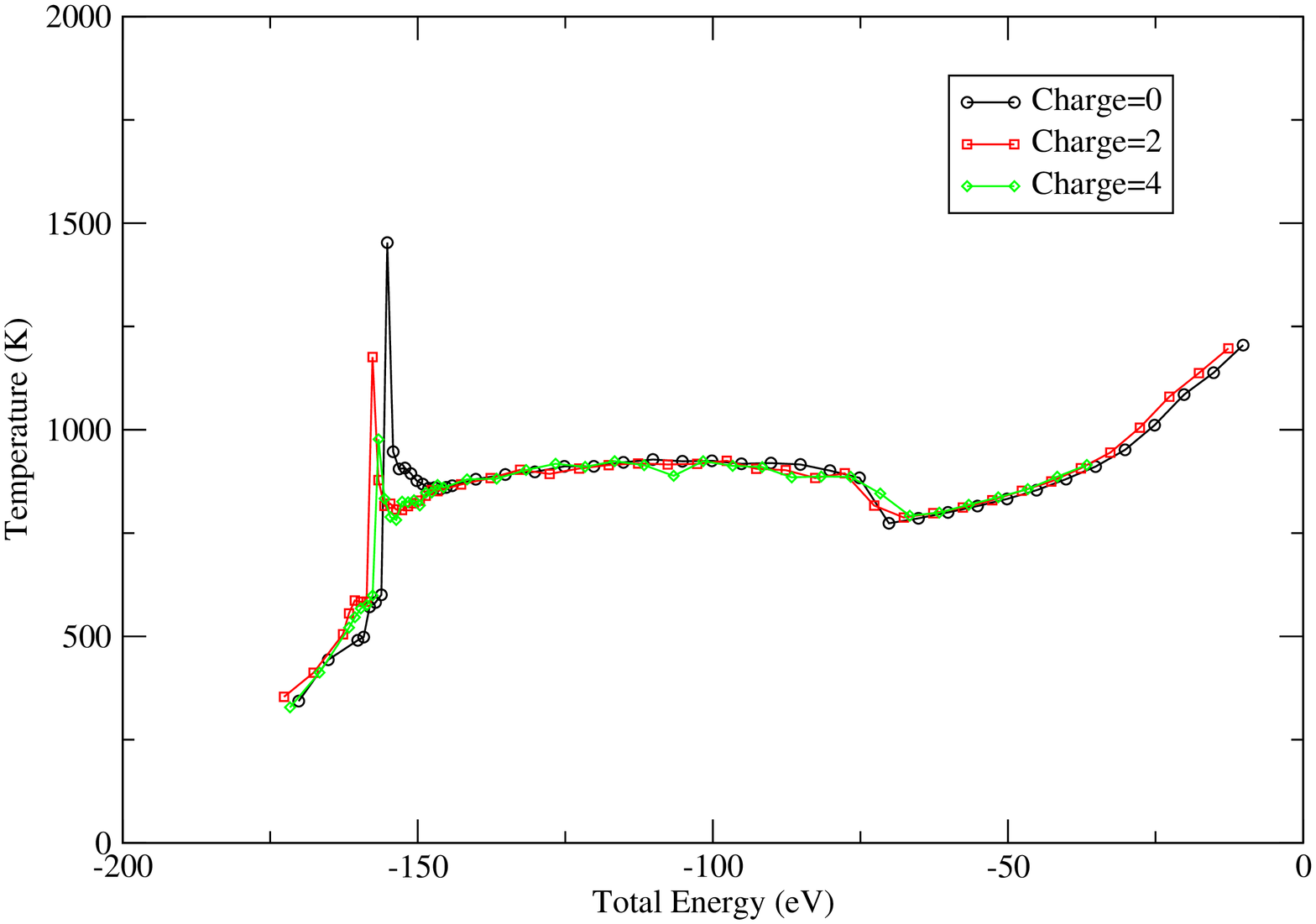}
\caption{Temperature curves for Na cluster of mass 200 atoms
and charges of 0, 2e, and 4e as a function of the total
energy of the system.\label{fig6}}
\end{figure}
\end{center}

In Figure \ref{fig6} we show the three caloric curves for Na 200 clusters 
for charges of 0, 2, and 4,
as a function of the total energy $E_{tot}=E+E_{bind}$. We note that there 
is not much
difference between the three curves: The plateau is the same, and the 
location of the
phase transition is the same for these cluster charges. The slope of the dip
becomes shallower as the charge of the cluster is increased.

\section{CONCLUSION}

In conclusion, we have calculated the mass distribution for the 
fragmentation of
sodium clusters of charges ranging
from 0 to 9. The calculations were carried out microcanonically and with an 
exact
treatment of detailed balance. The caloric curves were found to be 
independent of the charge on the cluster
for charges up to 4. Although it is not shown in the calculation presented 
here, the
caloric curve is somewhat sensitive to the radius parameter. Increasing the 
volume of the system
lowers the phase transition temperature, and shifts the dip at the end of 
the
plateau to lower excitation energy since the pressure is decreased, and 
smaller
volume favors the liquid or solid state. An increase in the system radius 
parameter
from $4\AA$ to $8\AA$ lowers the temperature at the plateau by about $200K$.
The calculation of the caloric curve
was unstable for clusters of mass 200 atoms and charges $\geq 5$, and this 
may be related to
the experimentally confirmed instability \cite{chand1,chand2,cguet} of such 
highly charged clusters which immediately decay by
emitting light singly charged fragments and cannot therefore achieve thermal 
equilibration which is the fundamental assumption of our approach.

\vspace{1cm}
\noindent Acknowledgments: This work is supported by the Deutsche 
Forschungsgemeinschaft grant GR 398/6-2. We are especially grateful
to O.Fliegans and Th.Klotz for improved the numerical method for 
respecting detailed balance in the individual Monte-Carlo moves.
This was previously done in MMMC only on average.

\end{document}